\documentclass[pdflatex,sn-mathphys-num]{sn-jnl}

\usepackage{graphicx}%
\usepackage{multirow}%
\usepackage{amsmath,amssymb,amsfonts}%
\usepackage{amsthm}%
\usepackage{mathrsfs}%
\usepackage[title]{appendix}%
\usepackage{xcolor}%
\usepackage{textcomp}%
\usepackage{manyfoot}%
\usepackage{booktabs}%
\usepackage{algorithm}%
\usepackage{algorithmicx}%
\usepackage{algpseudocode}%
\usepackage{listings}%
\usepackage[authormarkup=none]{changes}
\usepackage{url,hyperref,lineno,microtype}
\usepackage{booktabs}
\usepackage{multirow}
\usepackage{makecell}
\usepackage{comment}

\theoremstyle{thmstyleone}%
\theoremstyle{thmstyletwo}%

\theoremstyle{thmstylethree}%

\begin{document}
\title[Resolution-adaptive U-Net for MS lesion segmentation]{Learning from spatially inhomogenous data: resolution-adaptive convolutions for multiple sclerosis lesion segmentation}

\author*[1,3]{\fnm{Ivan} \sur{Diaz}}\email{ivandiazzz@gmail.com}

\author[1]{\fnm{Florin} \sur{Scherer}}

\author[1]{\fnm{Yanik} \sur{Berli}}

\author[1,3]{\fnm{Roland} \sur{Wiest}}

\author[2]{\fnm{Helly} \sur{Hammer}}

\author[2]{\fnm{Robert} \sur{Hoepner}}

\author[2]{\fnm{Alejandro Leon} \sur{Betancourt}}

\author[1,3,4]{\fnm{Piotr} \sur{Radojewski}}
\equalcont{These authors contributed equally to this work.}

\author[1,3]{\fnm{Richard} \sur{McKinley}}
\equalcont{These authors contributed equally to this work.}

\affil[1]{\orgdiv{Support Center for Advanced Neuroimaging (SCAN), University Institute of Diagnostic and Interventional Neuroradiology, , Bern, Switzerland}, \orgname{Inselspital, Bern University Hospital and University of Bern}, \orgaddress{\city{Bern}, \country{Switzerland}}}

\affil[2]{\orgdiv{Department of Neurology }, \orgname{Inselspital, Bern University Hospital and University of Bern}, \orgaddress{ \city{Bern}, \country{Switzerland}}}

\affil[3]{\orgdiv{Centre for Artificial Intelligence in Medicine}, \orgname{University of Bern}, \orgaddress{\city{Bern},  \country{Switzerland}}}



\abstract{
In the setting of clinical imaging, differences in between vendors, hospitals and sequences can yield highly inhomogeneous imaging data. In MRI in particular, voxel dimension, slice spacing and acquisition plane can vary substantially. For clinical applications, therefore, algorithms must be trained to handle data with various voxel resolutions.   The usual strategy to deal with heterogeneity of resolution is harmonization: resampling imaging data to a common (usually isovoxel) resolution. This can lead to loss of fidelity arising from interpolation artifacts out-of-plane and downsampling in-plane. We present in this paper a network architecture designed to be able to learn directly from spatially heterogeneous data, without resampling: a segmentation network based on the e3nn framework that leverages a spherical harmonic, rather than voxel-grid, parameterization of convolutional kernels, with a fixed physical radius. Networks based on these kernels can be resampled to their input voxel dimensions. We trained and tested our network on a publicly available dataset assembled from three centres, and on an in-house dataset of Multiple Sclerosis cases with a high degree of spatial inhomogeneity.  We compared our approach to a standard U-Net with two strategies for handling inhomogeneous data: training directly on the data without resampling, and resampling to a common resolution of 1mm isovoxels.  We show that our network is able to learn from various combinations of voxel sizes and outperforms classical U-Nets on 2D testing cases and most 3D testing cases.  This shows an ability to generalize well when tested on image resolutions not seen during training. Our code can be found at: \url{http://github.com/SCAN-NRAD/e3nn\_U-Net}.}


\maketitle

\section{Introduction} 

Deep-learning-based labeling and quantification of medical images stands poised to transform radiological practice, streamlining workflows and allowing precise, repeatable measurements. For algorithms to be applicable and unbiased in a clinical setting, they must be trained on data coming from a broad population of patients, drawing from diverse data sources, beyond that available in univesity hospital settings.  As an illustrative example, in multiple sclerosis (MS) Fluid-Attenuated Inversion Recovery MR sequence (FLAIR)  is essential for identifying white-matter lesions. A patient might be scanned at high-resolution (0.5x0.5x0.5 mm or better) in a highly specialised facility, with standard 3D imaging (1 mm$^3$ isovoxel), or with lower resolution 2D imaging (e.g. 1x1x3 mm). Furthermore, 2D imaging may not always be acquired in the same direction; the 2D slices may be in the coronal, sagittal or axial planes.   The standard practice for training networks on spatially heterogeneous datasets, as adopted for example in the nnU-Net framework~\cite{isensee_nnu-net_2021}, is to resample the imaging volumes to a common spatial resolution.  While this practice is well-established, it leads to a loss of fidelity when downsampling, and creates interpolation artifacts (with accompanying loss of real world performance) when upsampling.  

In this paper we propose an approach to learning from spatially heterogeneous data which does not require resampling to a common resolution, instead learning from minimally processed data in its native resolution. Our proposed method is a modified U-Net architecture~\cite{ronneberger2015u} in which the convolutions kernels are parametrized in physical space, as the sum of spherical harmonic basis function.  By defining the kernels in physical (rather than voxel) space, each convoluional layer can be realized at arbitrary spatial resolution.  The resulting \emph{Resolution Adaptive} convolutions can therefore be applied to, and learn from, images of any spatial resolution. 

Taking MS lesion segmentation as a motivating example, we assembled a spatially heterogeneous dataset of FLAIR imaging together with lesion segmentations. We then investigate three approaches to learning a lesion-segmentation from this data: our resolution adaptive network and two baseline networks, a U-Net trained directly on the minimally processed data, a U-Net and a U-Net trained on resampled data at a common 1mm isovoxel resolution.

\subsection{Related Works}

The nnU-Net framework ~\cite{isensee_nnu-net_2021} employs a resampling strategy, using the median voxel spacings found in the training data as a target resolution.   While this is a principled choice, it does not take into account any unknown testing resolutions.  Similarly, splatting ~\cite{brudfors2022fitting} has been successfully employed for end-to-end training of a network to segment MRI volumes of different resolutions: this method projects images into a \emph{mean space} before training on a regular U-Net, after which predictions are projected back into native space. In addition to relying on prior knowledge of testing resolutions, the splatting operation is computationally expensive.
The recently proposed HyperSpace~\cite{Jou_HyperSpace_MICCAI2024} proposes hypernetworks as a solution to spatial heterogeneity: in this framework an auxiliary network taking voxel dimensions as input is used to generate the weights of a U-Net operating on images of that resolution.  Training these networks requires heavy data augmentation to ensure that the hypernetwork is not overfit to existing image resolutions, and the resulting networks were outperformed on most tasks by a simple resampling benchmark.

\section{Materials and Methods}

\subsection{Resolution adaptive convolutions} 

The e3nn library~\cite{e3nn_arXiv,e3nn}  is a geometric deep learning framework providing primitives for rotation-, translation- and inversion- equivariant neural networks.  While many of its functions are diesigned to work on point cloud data or graphs, the voxel convolution function~\cite{weiler_3d} defines convolutional feature extractors whose kernels are defined by a function
\begin{equation}
	W(\vec{r}) = R(\vec{r})Y_l^m(\hat{r})
\end{equation}
where the \emph{radial function} $R(\vec{r})$ depends only on the magnitude of $\vec{r}$ and is restricted to be zero beyond a certain radius (the \emph{width} of the kernel), and the \emph{spherical harmonic function} $Y_l^m(\hat{r})$ depends only on the orientation of $\vec{r}$.  To yield a kernel consisting of the usual arrays of floating point numbers, in order to apply to voxel grid data, the kernel at the centers of the grid-points of the kernel. We have previously used this framework to define rotationally-equivariant 3D U-nets ~\citep{diaz2023end}, demonstrating that these networks are more data efficient than standard U-Nets, while obviating the need for rotational data augmentation. 

In the usual formulation, the vector input $\vec{r}$ to the kernel function has abstract dimensions measured in number of voxels, and the voxel grid is assumed to be isotropic. In this extension of that work, we consider the function $W$ to be the continuous analog of a classic convolution kernel (such as, for example, a Gaussian blur) operating in physical space, and the vector $\vec{r}$ to be measured physical units (in, say millimeters), not voxels. This allows the kernel to be realized at any resolution grid resolution: for example, a kernel with wigth 5mm may be realized as a 5x5x5 kernel in a 1mm isovoxel space, but a 11x11x1 kernel in a 0.5x0.5x3mm voxel space.  (see Fig~\ref{table:kernel_shape} for more examples) . Note that the learnable network parameters are not dependent on the size/shape of the convolution kernel in voxels, but a determined only by the parameterization of the radial function and the number of spherical harmonic basis elements allowed in the convolution.   Since the operation of realizing the kernel in voxel space is differentiable, the network can learn a common set of spherical harmonic kernels from multiple spatial resolutions without resampling.

\begin{table}[h]
\begingroup
\renewcommand{\arraystretch}{1.5} 
\setlength{\tabcolsep}{4.5pt} 
\begin{tabular}{c c c c c||c c c c }
                                            & \multicolumn{4}{c||}{\bf{1 mm isovoxel}} & \multicolumn{4}{c}{\bf{0.5x0.5x3 mm}} \\
\multicolumn{1}{c|}{\begin{tabular}[c]{@{}c@{}}\bf{U-Net level}\end{tabular}}      & 0 & 1 & 2 & 3 & 0 & 1 & 2 & 3 \\
\multicolumn{1}{c|}{\begin{tabular}[c]{@{}c@{}}\bf{Voxel grid} \\ \bf{dimensions (mm)}\end{tabular}} & 1,1,1  & 2,2,2 & 4,4,4 &  8,8,8  & 0.5,0.5,3 & 2,2,3 & 4,4,3 & 8,8,6 \\  \hline
\multicolumn{1}{c|}{\begin{tabular}[c]{@{}c@{}}\bf{Convolutional Kernel}\\ \bf{width (mm)} \end{tabular}} & 5mm       & 10mm     &  20mm      & 40mm       & 5mm     & 10mm  & 20mm    & 40mm       \\ 

\multicolumn{1}{c|}{\begin{tabular}[c]{@{}c@{}}\bf{Convolutional Kernel}\\ \bf{shape (voxels)} \end{tabular}} & 5,5,5 & 5,5,5 & 5,5,5 & 5,5,5 & 11,11,1 & 5,5,3 & 5,5,7       & 5,5,7       \\ \hline
\multicolumn{1}{c|}{\begin{tabular}[c]{@{}c@{}}\bf{Maxpooling Kernel}\\ \bf{width (mm)} \end{tabular}} & 2 mm       & 4mm     &  8mm      & 16mm       & 2mm     & 4mm  & 8mm    & 16mm       \\
\multicolumn{1}{c|}{\begin{tabular}[c]{@{}c@{}}\bf{Maxpooling Kernel}\\ \bf{shape (voxels)} \end{tabular}} & 2,2,2 & 2,2,2 & 2,2,2 & 2,2,2 & 4,4,1 & 2,2,1 & 2,2,2       & 2,2,2       \\ 
 \\

\end{tabular}
\endgroup
\caption{Voxel convolution and maxpooling kernel shapes and dimensions for the Resolution adaptive U-Net used in this paper, at the various downsampling depths, as realized for 1~mm  isovoxel and 0.5x~0.5x~3~mm input data}

\label{table:kernel_shape} 
\end{table}

\subsubsection{Resolution-adaptive U-Net architecture}

The resolution-adaptive U-Net used in this paper is derived by the original spherical harmonic U-Net formulation~\citep{diaz2023end} by replacing the existing spherical harmonic kernels with their resolution-adaptive counterparts, and also by introducing a resolution-adaptive form of maxpooling, designed to harmonize as much as possible the resolutions of feature maps at the deeper levels of the U-Net, selectively downsample in only each dimension so that the physical size of the downsampled voxel becomes more isotropic.

We define an isotropic physical pooling width for each maxpooling layer (in this paper, set at 2.0 mm for the first maxpooling layer and doubling at each subsequent layer). At each maxpooling step, the maxpooling kernel is defined to be the largest which fits inside this width, with a minimum pooling size of 1 in each direction.  Concretely, this means that if less than two voxels fit into the pooling width along a particular direction, then no pooling occurs in this direction.  An illustrative example of the behaviour of this kernel is shown in Table~\ref{table:kernel_shape}.

\subsubsection{Networks trained}

Our baseline model is a 3D U-Net modeled on the nnU-Net, with three maxpooling/upsampling steps. The network has two convolutional layers of 30 features before the first maxpooling, and the feature depth doubles after each maxpooling and halves after each upsampling. For the resolution-adaptive U-Net we follow our previous publication, learning 8 scalar features (dimension 1), 4 vector features (dimension 3) and 2 rank-2 tensor features (dimension 5) before the first maxpooling layer, with the radial function being represented as a sum of five basis functions. In the notation of the \texttt{e3nn} library, this combination of features is denoted \texttt{8x0e~+~4x1e~+~2x2e}, and has total scalar dimension of 30, matching the baseline U-Net. Similar to the standard U-Net, these number of features was doubled and halved at each downsampling and upsampling step, respectively. The kernel dimension of the resolution-adaptive model was 5 mm and the kernel size of the standard U-Net was 5 voxels, meaning that kernel shapes are identical between the regular and resolution-adaptive network on 1mm isovoxel data.

\subsubsection{Dataset}

\begin{figure}[!h] 
  \centering
  \includegraphics[width=0.9\textwidth]{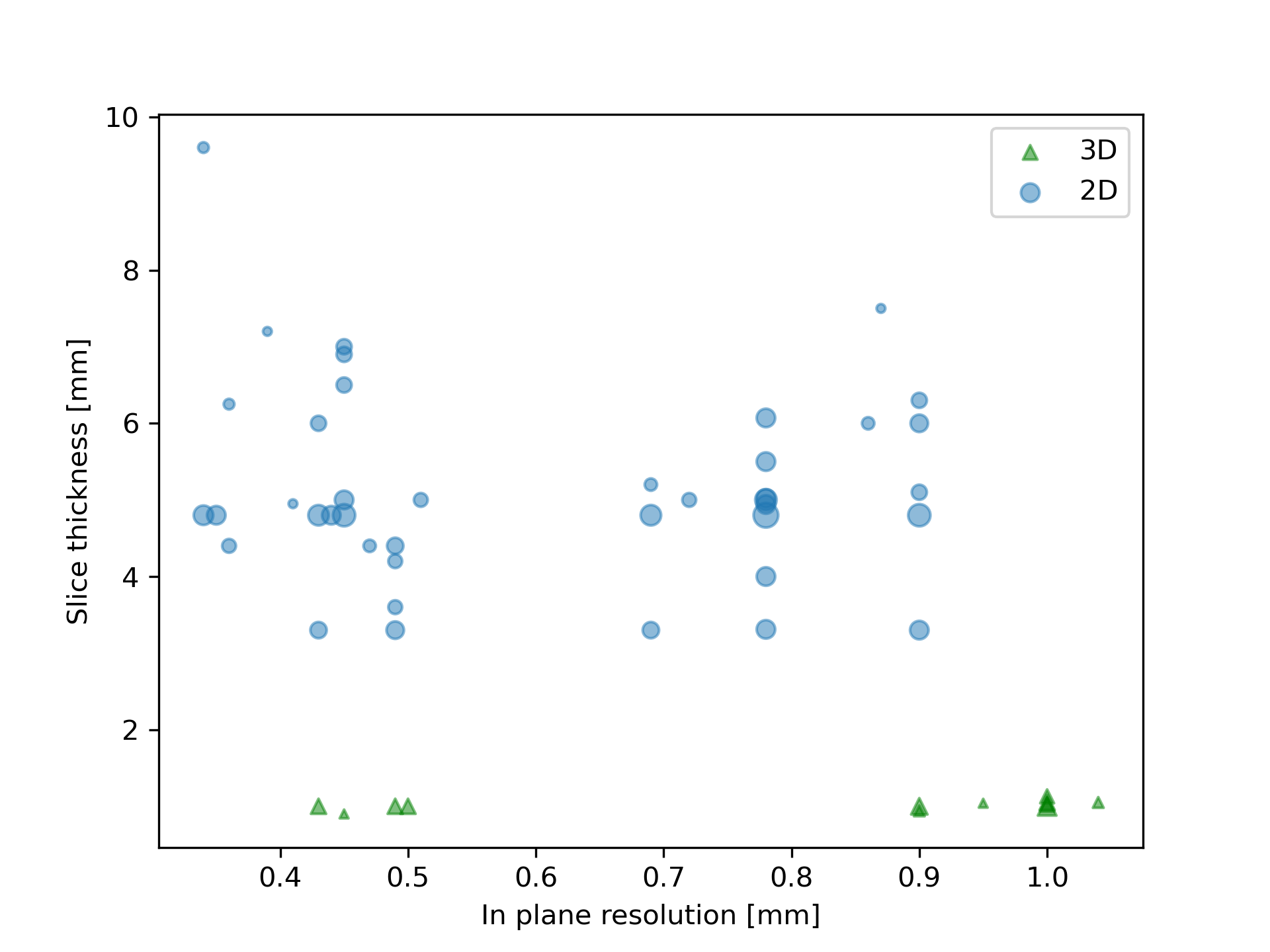}
    \caption{Distribution of 2D and 3D acquisitions from our in-house data set. The size of the markers is proportional to the number of cases available.} \label{voxel_dist}
\end{figure}

This study was approved by the respective ethic committee (registration no. KEK-BE ethic vote: Bern Neuroimmunological Registry: 2017–01369). Patient data identified from the existing neuroimmunological registry was used to assemble a dataset reflecting the clinical reality of MS imaging, including different images with varying scanners, field strengths, and resolutions. 
From the available data, 168 3D MRI and 172 2D MRI were segmented with an existing 2.5D segmentation model~\cite{mckinley2021simultaneous} and the results manually corrected.  Figure~\ref{voxel_dist} shows the distribution of resolutions for the 2D and 3D acquisitions. We see a wide variety of in-plane resolutions for both the 3D and 2D acquisitions.   

\subsection{Experiments} 

In all our experiments, we compare the performance of the resolution-adaptive and the ordinary 3D U-Net, with two different strategies for the ordinary U-Net: training on minimally processed data at its original resolution ("U-Net native") and  training on data resampled to 1mm isovoxel resolution by degree three bspline interpolation, with lesion masks resampled using nearest neighbor interpolation ("U-Net $@1\mathrm{mm}$").

\subsubsection{Experiments on 3D data}

The 3D data is primarily clustered around two resolutions: 3D isovoxel, and 3D slices with a 0.5mm in-plane resolution and 1mm slice thickness. Allowing for a deviation of up to 0.04 mm for cluster membership (so that a 0.49 mm and 1.02 mm in-plane volume would be included with the 0.5 mm and 1.0 mm in-plane clusters, respectively), we compare performance training and testing on data from these clusters.  Training data comprised 18 cases from the 0.5 mm in-plane cluster and 18 cases from the 1.0 mm in-plane cluster.   Testing data comprised 15 cases from the 0.5 mm in-plane cluster and 15 cases from the 1.0 mm in-plane cluster. Models were trained on each cluster separately and on the combined training data of both clusters, examining the extent to which models trained on one cluster can adapt to the other, and how well models can learn jointly from both clusters.  Our hypothesis is that the U-Net $@1\mathrm{mm}$ will perform less well in-sample (i.e. on previously seen resolutions), but better on unseen resolutions, than the U-Net native model, and that the Resolution-adaptive U-Net will perform better across both testing datasets.

\subsubsection{Experiments on 2D data and 3D}

To compare the ability of the three tested methods to learn from highly heterogeneous data, we retrained each model type combination of 0.5 mm in-plane cluster (68 cases), 1.0 mm in-plane cluster (68 cases) and 2D cases (68 cases). We tested these methods on the previous 3D tests sets as well as a set of 29 2D cases. 

\subsubsection{Model Training hyperparameters}

Networks were trained on 120x120x120 voxel patches with a batch size of one, using a soft dice loss function, with an Adam optimizer, a learning rate of $5\text{e-}3$
 and early stopping on the validation loss with a patience of 300 epochs.
 Model predictions were performed using patch-wise prediction 
with overlapping patches and Gaussian weighting \cite{isensee_nnu-net_2021}.  

\subsubsection{Model evaluation and Statistical testing}
In all cases, we used the Dice similarity metric to compare the segmentation output of the network to the manually-annotated ground truth.
To assess any differences between the Dice Coefficients segmentation methods, we employed appropriate nonparametric statistical tests (Wilcoxon signed rank test, as implemented in \texttt{scipy.stats}), with Bonferroni correction to account for multiple comparisons.

\section{Results}

\subsection{Training and evaluation on 3D data}
\begin{table}[h]
\begingroup
\renewcommand{\arraystretch}{1.2} 
\setlength{\tabcolsep}{8pt} 

\begin{tabular}{ccccc}

\bf{Training} &  &  & \textbf{Test set} &  \\ 
\bf{Dataset} & \bf{Model} & \bf{1.0 mm } & \bf{0.5 mm} & \bf{Combined 3D} \\ \toprule
& U-Net native& $0.84 \ (0.09)$ & $0.54 \ (0.08)$ & $0.69 \ (0.17)$ \\
1 mm& U-Net $@1\mathrm{mm}$ & $0.80 \ (0.09)$ & $0.75 \ (0.07) $ & $0.78 \ (0.08)$ \\
& Resolution adaptive  & $0.85 \ (0.07)$ & $0.78 \ (0.06)$ & $0.82 \ (0.07)$ \\\midrule

 & U-Net native & $0.70 \ (0.15)$ & $0.78 \ (0.08)$ & $0.74 \ (0.15)$ \\
0.5 mm & U-Net $@1\mathrm{mm}$ & $0.79 \ (0.11) $ & $0.73 \ (0.08) $ & $0.76 \ (0.10) $ \\
& Resolution adaptive & $0.75 \ (0.08)$ & $0.77 \ (0.08)$ & $0.76\ (0.08)$ \\ \midrule

 & U-Net native & $0.84 \  (0.08)$ & $0.81 \  (0.09) $ & $0.83 \ (0.09)$ \\
Both& U-Net $@1\mathrm{mm}$ & $0.83 \ (0.08)$ & $0.76 \ (0.09)$ & $0.79 \ (0.09)$ \\
& Resolution adaptive & $\mathbf{0.86 \ (0.06)}$ & $\mathbf{0.81 \ (0.08)} $ & $\mathbf{0.84 \ (0.08)}$ \\
\botrule\\

\end{tabular}
\caption{Mean (standard deviation) Dice coefficient for models trained and tested on 3D data (1mm in-plane data, 0.5mm in plane data, and both).  In each column the best performing method is indicated in bold.}
\endgroup
\label{table:02} 
\end{table}
Results of training and testing on 3D MRI data of differing in-plane resolutions is shown in Table~\ref{table:02}. For the two single-resolution training sets the results follow the anticipated pattern: the network operating on native data without resampling performs better in-sample than the network trained on resampled data, but suffers a drop in performance when applied to out-of-sample data with an unseen resolution.  In all cases, the difference between training on native resolution and resampled data was statistically significant: notably, the difference was even observed when training and testing on data already close to 1mm isovoxel resolution.  Meanwhile, the resolution adaptive network performed well on data of both seen and unseen resolution.  Furthermore, for each testing set the Resolution adaptive U-Net based on e3nn kernels and trained on mixed resolution data was the best-performing model: however differences were small and statistical significance ) did not survive correction for multiple comparisons.  

\subsection{Evaluation of 3D-trained models on 2D data}

As expected, the model trained and evaluated on resampled data 3D performed best when applied to resampled 2D data, with the model trained on mixed resolution achieving a mean Dice coefficient of 0.74.  Neither the native resolution nor resolution adaptive networks were able to generalize, achieving mean Dice coefficients of 0.38 and 0.39 respectively.

\subsection{Training and evaluation on combined 2D and 3D data}

\begin{figure}

\centering    
\includegraphics[width=.3\linewidth]{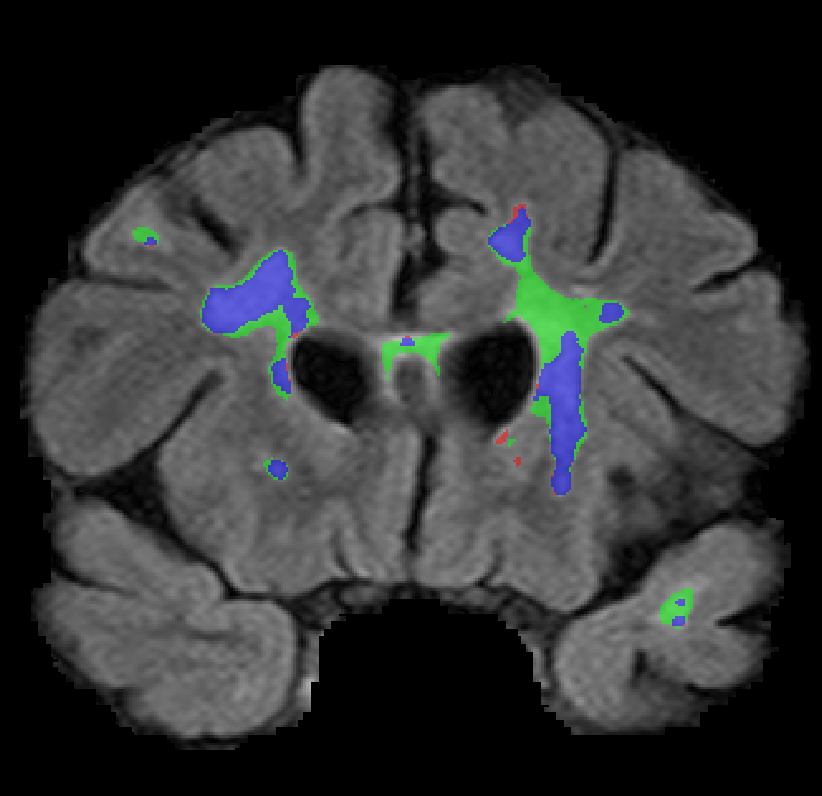}
\includegraphics[width=.3\linewidth]{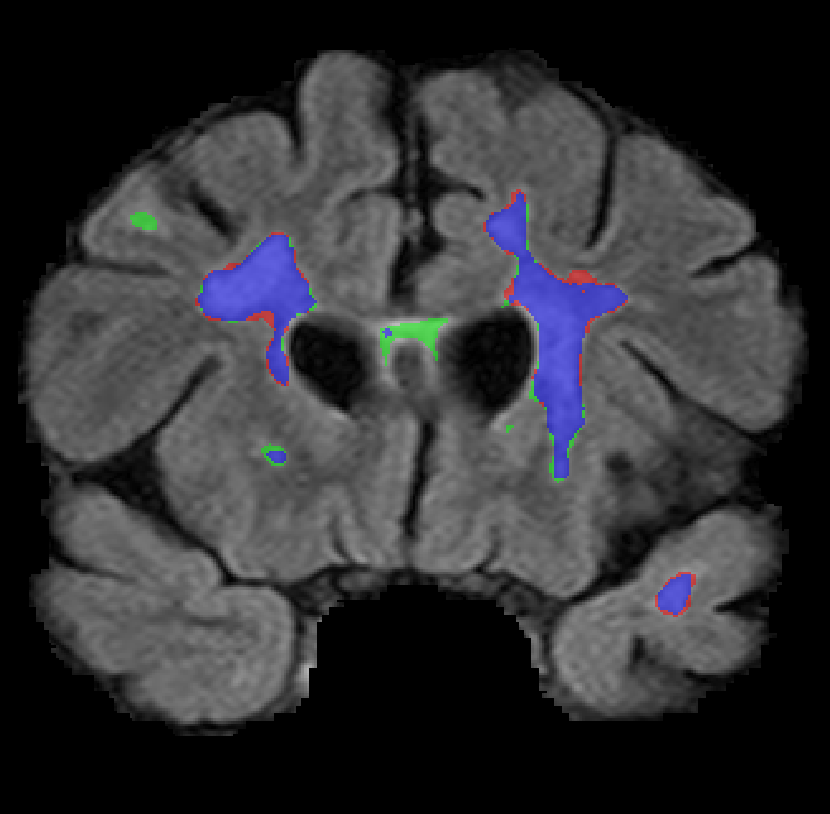}
\includegraphics[width=.3\linewidth]{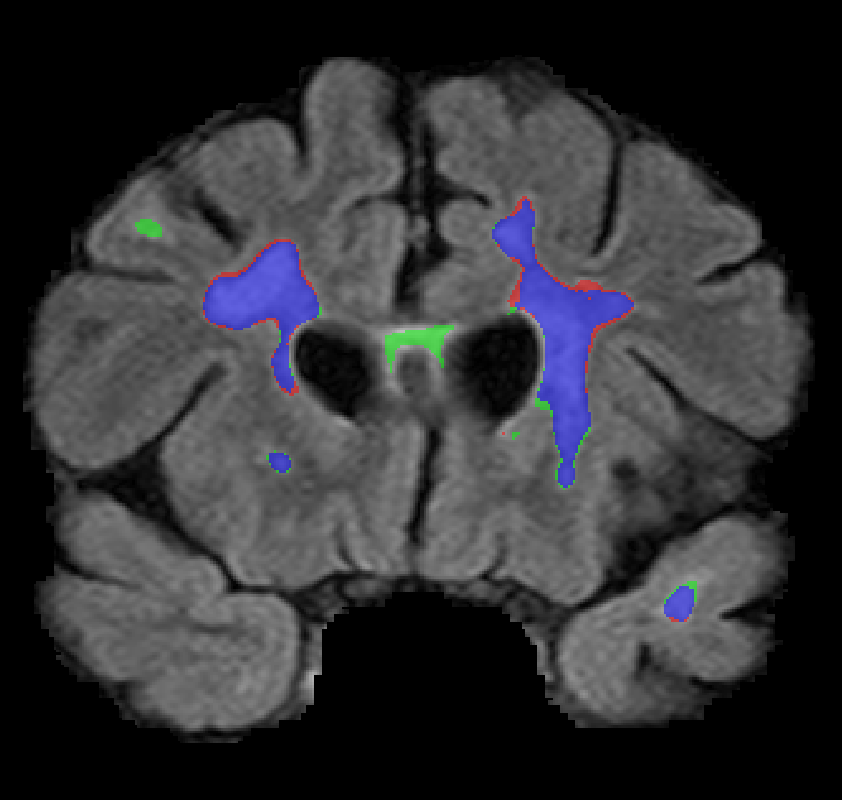}
\caption{Comparison of segmentations of a 2D FLAIR image (coronal acquisition, 0.35mm in-plane resolution, slice spacing 4.8mm), using models trained on both 3D and 2D data. Blue pixels denote a true positive, green are false negatives and red false positives. (a) U-Net native, (b) U-Net $@1\mathrm{mm}$ (c) Resolution-adaptive network.  The network operating at native resolution fails to segment large, areas at the center of the lesion, while the network trained and applied to resampled data performs better but cannot take advantage of the very high in-plane resolution, which allows the Resolution-adaptive network to better follow the lesion contours.  }
\label{fig:comparison}
\end{figure}

Results of the models trained and evaluated on the mixed dataset of 3D and 2D data are shown in Table~\ref{table:3D_2D}.  While for the 3D datasets the U-Net operating on native resolution data performed best, the difference to the resolution adaptive model was not significant.  However, for the testing dataset overall, and for the 2D testing cases, the resolution-adaptive method was superior, with statistically significant improvements surviving multiple comparisons correction.  An illustrative example showing segmentation performance on a 2D example is shown in Figure~\ref{fig:comparison}.

\begin{table}[!h]
\centering
\begingroup
\setlength{\tabcolsep}{5pt} 
\renewcommand{\arraystretch}{1.5} 
\begin{tabular}{c c c c c} 
\bf{Model} & \bf{1.0 mm in-plane} & \bf{0.5 mm in-plane} & \bf{2D} & \bf{2D and 3D}\\ \toprule

U-Net native& $\mathbf{0.89 \ (0.06)}$ & $\mathbf{0.84 \ (0.08)} $ & $0.77 \ (0.13)$ & $0.82 \ (0.11)$\\
U-Net $@1\mathrm{mm}$ & $0.86 \ (0.07)$ & $0.82 \ (0.06)$ & $0.74 \ (0.14)$ & $0.82 \ (0.08)$\\
Resolution Adaptive & $0.89 \ (0.06)$ & $0.83 \ (0.07)$ & $\mathbf{0.82^{*} \ (0.06)}$ & $\mathbf{0.84^{*} \ (0.08)}$\\
\botrule\\

\end{tabular}
\endgroup
\caption{Mean (standard deviation) of Dice coefficient for models trained on both 3D (1mm and 0.5mm in plane) and 2D volumes.  In each column, the best performing method is indicated in bold, with an asterisk if it was found significantly better than all other methods in that column (Wilcoxon signed rank test, 5\% significance level, Bonferroni-corrected)}
\label{table:3D_2D} 
\end{table} 

\section{Discussion}

Our experiments at 0.5mm and 1mm in-plane resolution consistently show that U-Net models trained on data at their original resolution perform better in-sample than their resampled counterparts, with very similar performance between the baseline U-Net and the resolution adaptive model.  Out-of-sample, the network trained on native resolution images performed poorly: we observe a similar pattern as seen in HyperSpace~\cite{Jou_HyperSpace_MICCAI2024}, with a reduction in voxel size leading to a larger drop in performance than in increase in voxel size.  This pattern was not observed with the resolution-adaptive network, which substantially outperformed the baseline U-Net operating on native data, suggesting that the harmonic U-Net kernels were less strongly overfit to their training data. When both resolution clusters were available as training data, the resolution adaptive U-Net appears better able to learn from the combined data.  

It is striking that the worst-performing network in the 3D experiments was trained and tested on resampled data.  This network only outperformed on 2D data which is substantially out-of-sample. When 2D training data was also available, however, both the native space U-Net and the resolution adaptive network outperform training on resampled data.  This suggests that resampling is only advisable in circumstances where testing data may come from substantially different sequences as those available during training. 

In this paper, we have focused on the question of MS lesion segmentation, owing to the availability at our institution of highly diverse data.  The majority of available datasets in medical imaging do not exhibit this variability, either because they are based on high-quality research data, or because they are made available already in a resampled format.  The conclusions of this study cannot yet be assumed to apply in all domains, and there could be settings where training on resampled data is superior to training on native data, for example where structures are large and have ill-defined borders. 

The increased performance of the resolution adaptive network comes at a cost of computationally more intensive training, mainly owing to the need to separately instantiate different instances of the U-Net at different resolutions.  When multiple resolutions are present for training, this leads to an increase in time between batches if training is performed on a single GPU, which can of course be circumvented at the cost of training over multiple GPUs. Apart from this,  training dynamics are similar to those of a standard network, with training being relatively insensitive to initializations and learning rates, which is often not the case for hypernetworks~\cite{Jou_HyperSpace_MICCAI2024, chauhan2024}. A planned reimplementation of the primitives of the e3nn package, on which our network relies, in the jax framework, may in the future yield resolution adaptive networks which are both easier to train and more performative.

\section{Conclusion} 

Training machine learning models on real-world clinical datasets makes it necessary to deal with spatially heterogeneous data. The standard approach to dealing with heterogeneity is resampling to a common image resolution: as we have shown in this paper this buys robustness to varying image resolution at the price of significantly reduced in-sample performance. Our results suggest that for mixed resolution 3D training data it may be preferable to train at native resolution, while for more varied data resampling is required.  The decision of which approach to take for a given dataset is empirical. Our proposed resolution-adaptive model, based on a spherical-harmonic representation of convolutional kernels, offers an alternative to that choice which works well in both settings. Code to implement and train these networks is available at \url{https://github.com/SCAN-NRAD/e3nn_U-Net}.

\subsubsection{Acknowledgements} This work was supported by Spark Grant CRSK-3\_195801 and by the Research Fund of the Center for Artificial Intelligence in Medicine, University of Bern, for 2022-23 and Schweizerische Herzstiftung grant no FF10859.

 \bibliography{refs}

\end{document}